**Projected hydrologic changes over the north of the Iberian Peninsula using a Euro-CORDEX multi-model ensemble**

Patricio Yeste[(1)], Juan José Rosa-Cánovas[(1)], Emilio Romero-Jiménez[(1)], Matilde García-Valdecasas Ojeda[(2)], Sonia R. Gámiz-Fortis[(1)], Yolanda Castro-Díez[(1)] and María Jesús Esteban-Parra[(1)]

**(1)** Dept. Applied Physics, University of Granada, Spain

**(2)** Instituto Nazionale di Oceanografia e di Geofisica Sperimentale (OGS), Italy

**Corresponding author**: Patricio Yeste (pyeste@ugr.es)

**Abstract**

This study explores the impacts of climate change on the hydrology of the headwater areas of the Duero River Basin, the largest basin of the Iberian Peninsula. To this end, an ensemble of 18 Euro-CORDEX model experiments was gathered for two periods, 1975-2005 and 2021-2100, under two Representative Concentration Pathways (RCP4.5 and RCP8.5), and were used as the meteorological forcings of the Variable Infiltration Capacity (VIC) during the hydrological modelling exercise. The projected hydrologic changes for the future period were analyzed at annual and seasonal scales using several evaluation metrics, such as the delta changes of the atmospheric and land variables, the runoff and evapotranspiration ratios of the overall water balance, the snowmelt contribution to the total streamflow and the centroid position for the daily hydrograph of the average hydrologic year. Annual streamflow reductions of up to 40% were attained in various parts of the basin for the period 2071-2100 under the RCP8.5 scenario, and resulted from the precipitation decreases in the southern subwatersheds and the combined effect of the precipitation decreases and evapotranspiration increases in the north. The runoff and the evapotranspiration ratios evinced a tendency towards an evaporative regime in the north part of the basin and a strengthening of the evaporative response in the south. Seasonal streamflow changes were mostly negative and dependent on the season considered, with greater detriments in spring and summer, and less intense ones in autumn and winter. The snowmelt contribution to the total streamflow was strongly diminished with decreases reaching -80% in autumn and spring, thus pointing to a change in the snow regime for the Duero mountains. Finally, the annual and seasonal changes of the centroid position accounted for the shape changes of the hydrograph, constituting a measure of seasonality and reflecting high correlations degrees with the streamflow delta changes.

## 1. Introduction

Global water resources are expected to undergo vital changes as a consequence of the increasing temperatures and the varying precipitation regimes projected for the future climate (IPCC, 2014). The global water cycle is governed by the partitioning of precipitation into evapotranspiration and runoff (Saha et al., 2020), and despite the importance of the future changes in precipitation, the changes in evapotranspiration and runoff can play an even more meaningful role for the assessment of future water security (Lehner et al., 2019). The vulnerability of the hydrological cycle to those changes constitutes a major concern and a key challenge for the hydrologic community (Clark et al., 2016; Blöschl et al., 2019a; Nathan et al., 2019), urging the need to develop effective adaptation strategies that mitigate the different hydrologic stresses (Clark et al., 2016; Garrote et al., 2016). Particularly, climate change is likely to increase the frequency and intensity of hydrologic extremes (Blöschl et al., 2019a; Yang et al., 2019), such as droughts (Tomas-Burguera et al., 2020) and floods (Vormoor et al., 2015), as well as the alteration of the freshwater availability and the snow dynamics in mountainous systems (Viviroli et al., 2011; Mankin et al., 2015; Thackeray et al., 2019).

Although many efforts have been made to identify the emergence of the climate change signal under a wide range of climate scenarios (Taylor et al., 2012; O'Neill et al., 2016; Lehner et al., 2019), its effects are already evident in certain regions and are expected to become stronger with the increase of greenhouse gas (GHG) emissions (IPCC, 2014). This is the case of the Mediterranean Basin, where water scarcity and the occurrence of extreme events have strengthened over the 20th century (García-Ruiz et al., 2011; Garrote et al., 2016). For instance, Tramblay et al. (2020) indicate a growing frequency and severity for Mediterranean droughts. Floods have shown a downward trend in many catchments of southern Europe over the last decades (Blöschl et al., 2019b, Tramblay et al., 2019), presumably due to decreasing precipitation and increasing evaporation ratios, and resulting in diminutions of up to 23% per decade (Blöschl et al., 2019b). However, catchments belonging to north-western Europe have manifested increasing floods of about 11% per decade (Blöschl et al., 2019b). This tendency is also noticeable for small catchments of few squared kilometers in south-western Europe (Amponsah et al., 2018), where enhanced convective storms and land-cover changes may cause flash floods to increase (Blöschl et al., 2019b).

Approximately one half of the water scarcity areas of the Mediterranean Basin are located in southern Europe (Iglesias et al., 2007; Garrote et al., 2016), where the runoff reductions can present a threat for meeting the water supply needs of the agricultural, industrial and urban water demands (García-Ruiz et al., 2011). Notably, the south-western sector of the Mediterranean region, represented by the Iberian Peninsula, has been identified as a hotspot particularly vulnerable to the climate change impacts (Diffenbaugh and Giorgi, 2012; Marx et al., 2018; García-Valdecasas Ojeda et al., 2020a, 2021). Precipitation is expected to decrease over this region under climate change scenarios, with marked projected reductions in autumn, spring and summer for Spain (Argüeso et al., 2012; García-Valdecasas Ojeda et al., 2020a, b) and Portugal (Soares et al., 2017). Projected evapotranspiration changes over the Iberian Peninsula reflect considerable spatio-temporal variability (García-Valdecasas Ojeda et al., 2020a, b) and result from the interplay between the future water availability in the soil and a higher atmospheric demand driven by increasing temperatures (Jerez et al., 2012), with a trend towards soil-drying conditions by the end of the 21st century (García-Valdecasas Ojeda et al., 2020a). On the other hand, there is substantial evidence for the decrease of Iberian streamflows during the last half of the 20th century (Lorenzo-Lacruz et al., 2012), with a strong consensus about this trend for different Iberian catchments under present and future climate conditions (Salmoral et al., 2015; Gampe et al., 2016; Pellicer-Martínez and Martínez-Paz, 2018; Yeste et al., 2018; Fonseca and Santos, 2019).

Climate change impact studies are mainly based on the analysis and application of the projections carried out with General Circulation Models (GCMs) and Regional Climate Models (RCMs) (Pastén-Zapata et al., 2020). While the conceptualization of the Earth System processes is common for both modelling approaches, their primary difference lies in the spatial resolution of the implemented domain, typically set at 2.5º for GCMs (Tapiador et al., 2020) and allowing a more accurate representation of the regional and local characteristics in the case of RCMs (Rummukainen, 2010; Teutschbein and Seibert., 2010). Nonetheless, the increasing computing power has led to a progressive refinement of the spatial resolution of GCMs, sometimes exceeding and improving the RCMs resolution, and are expected to be superseded by high-resolution GCMs in the next generation of climate model simulations (Tapiador et al., 2020). Anyhow, RCMs projections still remain as a valuable and suitable data source for impact studies given the lack of widespread availability of high-resolution GCM projections, and constitute an appropriate tool for the evaluation of hydrologic changes at the basin scale (Pastén-Zapata et al., 2020). In this respect, the Euro-CORDEX project (Jacob et al., 2014) is established as the largest climate modelling effort for the European region (Herrera et al., 2020), with a plethora of RCM simulations available at 0.11º and 0.44º that have been the basis of a great number of hydrological impact studies for many European catchments (e.g. Gampe et al., 2016; Papadimitriou et al., 2016; Meresa and Romanowicz, 2017; Hakala et al., 2018; Hanzer et al., 2018; Vieira et al., 2018; Fonseca and Santos, 2019; Pastén-Zapata et al., 2020).

Using a Euro-CORDEX multi-model ensemble, this study aims to identify and analyse the projected hydrologic changes for an important basin located in the north of the Iberian Peninsula, the Duero River Basin. The Duero basin has been previously studied mainly from a statistical perspective focused on various hydroclimatic and land-surface variables under present climate. For instance, Ceballos-Barbancho et al. (2008) and Morán-Tejeda et al. (2010) reported the impacts of land-cover changes on water availability and water resources management for the basin during the last half of the 20$^{th}$ century. Morán-Tejeda et al. (2011) provided useful insights on the different river regimes characterizing the Duero and its tributaries. More recently, Fonseca and Santos (2019) studied the impacts of climate change for the Tâmega River, a northern tributary of the Duero River located in Portugal, using a Euro-CORDEX ensemble as well.

In this work, the Variable Infiltration Capacity (VIC) model (Liang et al., 1994; 1996) has been implemented for various headwater subwatersheds of the Duero basin based on the previous calibration exercise carried out in Yeste et al. (2020) for the study area. The VIC model largely improved the benchmark performance against streamflow observations and two actual evapotranspiration products, ensuring the further applicability of the calibrated parameters for the modelling exercise here developed. The main objective of this work consists of evaluating the future changes of the different hydrologic variables involved in the water balance at annual and seasonal scales, adopting different interrelated approaches that accurately highlight many fundamental features of the future hydrologic behaviour of the basin.

**2. Study area**

The Duero River Basin is an international basin located in the north of Spain and Portugal and represents the largest basin of the Iberian Peninsula (98 073 km$^2$). The focus of this study is placed on the 80% of its area, corresponding to the Spanish territory (Fig. 1). The topography of the basin is mainly constituted by a large central depression and the surrounding mountain chains that configure the headwater areas of the hydrologic network. The mean annual precipitation volume is around 50 000 hm$^3$ and mostly evaporates into the atmosphere (~35 000 hm$^3$), representing the remaining volume the water contribution of the basin as natural runoff. With a predominant Mediterranean climate, most of the precipitation occurs in the mountainous systems, exceeding 1000 mm/year in the northern mountains and showing values below 1000 mm/year in the southern part of the basin. It is concentrated in the autumn, winter and spring months, with a dry period affecting the majority of the area during summer, with a warmer temperature (~20.5 ºC in July).

The selection of the subwatersheds for this study (Fig. 1, Table S1 in supplementary material) was based on the implementation of the VIC model carried out in Yeste et al. (2020) for the headwaters of the Duero River Basin. The Nash-Sutcliffe Efficiency (*NSE*, Nash and Sutcliffe, 1970) was selected there as the main evaluation metric, and despite the good performance for the majority of the studied subwatersheds, some of them showed poor *NSE* estimations. A threshold *NSE* value of 0.67 was set in this study as an acceptable model performance based on previous studies (e.g. Martinez and Gupta, 2010; Ritter and Muñoz-Carpena, 2013; Her et al., 2019). This criterion reduced the number of subwatersheds considered to 24 out of the 31 originally included in Yeste et al. (2020) (see Table S1).

## 3. Data and methods
### 3.1 Euro-CORDEX multi-model ensemble

Daily climate data were gathered from the Euro-CORDEX project (http://www.euro-cordex.net/) at a spatial resolution of 0.11º (EUR-11, ~12.5 km) for eight atmospheric variables: precipitation, maximum and minimum temperature, near-surface wind speed, incoming shortwave and longwave radiation, atmospheric pressure and vapour pressure. The dataset was regridded to 0.05º (~ 5 km) using the Climate Data Operator (CDO) software (Schulzweida, 2019) and choosing a nearest neighbour assignment for the subsequent hydrological modelling exercise. The multi-model ensemble consists of 18 RCM+GCM combinations for the Representative Concentration Pathways (RCPs) 4.5 and 8.5. The ensemble list is provided in Table 2. Data were extracted for the historical period of 1975-2005 and for 2021-2100 as the future period considering the hydrologic year (i.e. from October to September) and the associated hydrologic seasons. The latter was divided into three sub-periods for the analysis: short-term (2021-2050), mid-term (2041-2070) and long-term (2071-2100) future periods.

**3.2 Bias correction**

The straightforward application of raw RCM data for hydrological impact studies is inadequate given the emerging systematic errors (i.e. biases) during the dynamical downscaling of GCM outputs (Gudmundsson et al., 2012; Hanzer et al., 2018). These uncertainties are usually managed with the use of ensembles of RCM simulations and the application of bias correction techniques (Déqué, 2007; Teutschbein and Seibert, 2012). Within the different bias correction methods, the Quantile Mapping (QM) (e.g. Wood et al., 2002; Déqué, 2007; Themeßl et al., 2011) has shown to produce better results (Themeßl et al., 2011; Teutschbein and Seibert, 2012; Hakala et al., 2018) and allows the correction of daily precipitation and temperature data (Meresa and Romanowicz, 2017; Hakala et al., 2018; Pastén-Zapata et al., 2020).

In this study we used the R package 'qmap' (Gudmundsson et al., 2012) in order to fit the cumulative distribution functions (CDFs) of the meteorological time series to the CDFs of the observations. Precipitation, maximum temperature and minimum temperature were the only bias-corrected variables given the absence of observations for the rest of the meteorological fields. For this end, daily precipitation, maximum temperature and minimum temperature data were gathered from the observational datasets SPREAD (Serrano-Notivoli et al., 2017) and STEAD (Serrano-Notivoli et al., 2019) for the historical period. SPREAD and STEAD are gridded datasets that cover Peninsular Spain and the Canary and Balearic Island with a spatial resolution of 5 x 5 km, and were built using data from an extensive net of observatories (>12000 for SPREAD and >5000 for STEAD). The QM method was then applied for each month of the year using pooled daily data. In the case of daily temperature, the QM technique was applied to the diurnal temperature range (DTR) and the maximum daily temperature. This approach effectively avoids the occurrence of negative DTR values and improves the posterior estimation of minimum daily temperature (Thrasher et al., 2012).

**3.3 VIC model**

The VIC model (Liang et al., 1994, 1996) is a macroscale hydrologic model that solves the water and energy balances at a grid cell level. The most common implementation of the model consists of three soil layers (VIC-3L), where the runoff is generated via surface and subsurface processes. Surface runoff is generated as an infiltration excess from the top two soil layer following a variable infiltration curve (Zhao et al., 1980). An ARNO formulation (Francini and Pacciani, 1991) is applied to the bottom soil layer, and the baseflow component is divided in two parts: 1) a linear law at low soil moisture contents; and 2) a quadratic response for higher moisture contents. Potential evapotranspiration is calculated through the Penman-Monteith equation, and the actual evapotranspiration is derived from the sum of three components: evaporation from bare soil, evaporation from interception and transpiration. A snow model for the representation of the accumulation and melting processes is applied for each grid cell through the definition of snow bands, thus taking into account the sub-grid variability accompanying the large grid size.

The VIC model was previously calibrated in Yeste et al. (2020) for the headwaters subwatersheds of the Duero River Basin against streamflow observations, and its performance was evaluated for the streamflow and actual evapotranspiration simulations following a benchmark approach. The VIC model improved the benchmark performance in both cases, making it suitable for further applications using the calibrated parameters. A more detailed explanation of the soil and vegetation parameterizations and the aggregation method for the outputs of the model, together with an in-depth sensitivity analysis for the calibration parameters, are provided in Yeste et al. (2020).

## 3.4 Snowmelt contribution to the total streamflow

The total runoff simulated with the VIC model feeds on water arising from both rainfall and snowmelt that infiltrates into the soil. However, the proportion of runoff corresponding to each of them is not explicitly accounted for (Siderius et al., 2013), and can be calculated for each month as follows:

$$Q_{snow,i} = min\left\{Q_i \cdot \left[\frac{melt_i}{rain_i + melt_i}\right], melt_i\right\} (1)$$

For the month $i$, $Q_{snow,i}$ [L³/T] is the streamflow arising from the snowmelt (i.e. melt streamflow), $Q_i$ [L³/T] is the total streamflow, $melt_i$ [L³/T] is the snowmelt rate and $rain_i$ [L³/T] the rainfall rate. This approach is analogous to that applied in Siderius et al. (2013) and Li et al. (2019), and constitutes an appropriate method for the estimation of the snowmelt contribution to the total runoff. Eq. (1) assumes that $Q_{snow,i}$ cannot exceed the melt-to-rain ratio nor the total snowmelt occurring for a given month, thus not introducing an imbalance conducive to unrealistic values. Note that in this work we use the terms "runoff" and "streamflow" interchangeably due to the aggregation method applied in Yeste et al. (2020) to the raw gridded outputs from VIC for obtaining the hydrologic time series at the subwateshed scale.

## 3.5 Evaluation metrics and projected hydrologic changes

The projected hydrologic changes were firstly analysed applying the delta change approach (Hay et al., 2000) to the mean annual and seasonal values of precipitation (*P*), potential evapotranspiration (*PET*), actual evapotranspiration (*AET*), total streamflow (*Q*) and melt streamflow ($Q_{snow}$). The statistical significance of the delta-changes was evaluated using the Mann-Whitney U test at 95% confidence level.

The future changes were then evaluated using five hydrologic signature measures (Stewart et al., 2005; Rasmussen et al., 2014; Mendoza et al., 2015; Mendoza et al., 2016). The first two measures provide information about the overall water balance for a certain region and can be derived from the water balance equation normalized by *P*:

$$1 = \frac{Q}{P} + \frac{AET}{P} + \frac{\Delta S}{P} \quad (2)$$

Where Δ*S* is the variation of the storage in the hydrological system for a given period, and all the variables are expressed in units of volume. For long periods, the storage component can be neglected (Rasmussen et al., 2014; Mendoza et al., 2015; Mendoza et al., 2016), and Eq. (2) can be rewritten as:

$$1 = \frac{Q}{P} + \frac{AET}{P} = i_Q + i_E \quad (3)$$

Where $i_Q$ is the runoff ratio and $i_E$ is the evapotranspiration ratio. Eq. (3) represents the steady-state for the water balance, and implies that both measures are complementary. Therefore, $i_Q$ and $i_E$ will be used to evaluate the present and future partitioning of precipitation into runoff and actual evapotranspiration at annual scale.

The third signature measure is the snowmelt contribution ratio to the total streamflow, $Q_{snow}$ ratio, and can be calculated both at annual and seasonal scales as follows:

$$Q_{snow} ratio = \frac{Q_{snow}}{Q} \quad (4)$$

Lastly, the centroid position for the daily hydrograph of the average hydrologic year was selected for the diagnosis of the projected *Q* changes. The most common metric is the X coordinate of the centroid for the entire hydrologic year or "center time" of runoff (Stewart et al., 2005; Mendoza et al., 2015; Mendoza et al., 2016), and evaluates the seasonality of runoff. In this study we have calculated both the X and Y coordinates ($C_X$, $C_Y$) at annual and seasonal scales, as shown in Fig. 2. The annual and seasonal centroids together provide a more accurate picture about the daily hydrograph. In addition, $C_Y$ is a valuable source of information about the shape of the hydrograph, and its future changes are related to the annual and seasonal delta-changes of *Q*. The correlation between the $C_Y$ changes and the projected *Q* delta-changes was tested through a linear adjustment calculating the coefficient of correlation *r* and the corresponding slope and intercept values. In this case the statistical significance of the *r* values was calculated using the Student's-t test at 95% confidence level.

**3.6 Model validation**

The predictive capability of VIC was firstly tested through the *NSE* values corresponding to the monthly *Q* and *AET* ensemble simulations for the period Oct 2000 – Sep 2011. This period was selected based on the prior calibration exercise of Yeste et al. (2020), and spans the last part of the historical period and the first years of the two RCP scenarios (historical+RCP validation periods hereafter). The analysis was carried out for both historical+RCP validation periods. Similarly to Yeste et al. (2020), *Q* simulations were validated against monthly streamflow observations gathered from the Spanish Centre for Public Work Experimentation (CEDEX, *Centro de Estudios y Experimentación de Obras Públicas*), and *AET* was compared to the monthly outputs of the Global Land Evaporation Amsterdam Model (GLEAM) version 3.3a (Martens et al., 2017; Miralles et al., 2011).

Lastly, the suitability of VIC to simulate the evaluation metrics previously described was analyzed for the two historical+RCP validation periods. The annual partitioning of precipitation into runoff and evapotranspiration was only evaluated for $i_Q$ using streamflow observations and SPREAD precipitation, assuming the steady-state for the water balance represented in Eq. (3). The annual and seasonal $C_X$ and $C_Y$ values were compared to those obtained with daily streamflow observations. The validation for $Q_{snow}$ ratio was not possible given the limited amount of observations for snow-related variables in the Iberian Peninsula.

**4. Results**

**4.1 Validation results**

The *NSE* values of *Q* and *AET* for the historical+RCP8.5 validation period are shown in Fig. 3a and Fig. 3b, respectively, and are almost identical to those calculated for the historical+RCP4.5 validation period (Fig. S1a and Fig. S1b in the supplementary material). *NSE(Q)* values are highly underestimated using the Euro-CORDEX multi-model ensemble in comparison to the calibration results in Yeste et al. (2020). This indicates that, while being an appropriate measure for calibrating hydrologic models, *NSE(Q)* constitutes a high-end performance extremely difficult to achieve when using climate model outputs. The runoff performance is commonly checked for less demanding metrics such as $i_Q$, which is calculated from mean values of runoff and precipitation (Eq. (3)) and plays a major role in assessing future water security (Lehner et al., 2019). Contrarily to *NSE(Q)*, the $i_Q$ biases of the ensemble and the calibration presented similar distributions with most values falling in the ±0.1 range (Fig. 3c and Fig. S1c in the supplementary material).

On the other hand, the ensemble clearly improved the VIC performance for the *AET* simulations, presumably due to the model-nature of GLEAM. Further calibration efforts for the Duero basin will aim at improving the VIC performance for *AET* targeting both *Q* and *AET* simulations simultaneously.

Finally, the annual and seasonal $C_X$ and $C_Y$ biases for the historical+RCP8.5 validation period (Fig. 4) resemble the ones estimated for the historical+RCP4.5 validation period (Fig. S2 in the supplementary material). Overall, the CDFs corresponding to the ensemble suggest an acceptable performance when compared to the calibration results from Yeste et al. (2020). The ensemble reflected a higher presence of positive biases, while the calibration presented slightly steeper CDFs closer to 0. Notably, the $C_X$ biases at annual scale are comparable to those showed in Mendoza et al. (2015, 2016) for three headwater subwatersheds in the Colorado River Basin.

## 4.2 Annual delta changes of *P, PET, AET and Q*

The mean annual values of *P*, *PET* and *AET* for the historical period are depicted in Fig. 5a, and the mean annual values of *Q* for this period are gathered in Table 2. A marked latitudinal gradient was found for the atmospheric variables, with *P* values generally above 1000 mm/year and *PET* values below 1000 mm/year for the northern subwatersheds, and reaching minimum *P* and maximum *PET* in the south. *AET* shows a narrower range of variability with the majority of values falling between 400 and 700 mm/year. The latitudinal gradient is also noticeable for this variable and reflects an opposite spatial distribution to *PET*.

Fig. 6 collects the annual delta changes of *P, PET, AET* and *Q* for the period 2071-2100 under the RCP8.5 scenario. The changes corresponding to RCP4.5 for the three future periods, and to the RCP8.5 for the 2021-2050 and 2041-2070 are shown in supplementary material (Fig. S3 to Fig. S7). A generalized decrease of annual *P* is expected for all the future study periods presenting maximum decreases of up to 40% in the south for the period 2071-2100 under the RCP8.5 scenario. *PET* is subject to significant increases for all the future study periods and RCPs, with maximum increases, above 40%, taking place in the northern subwatersheds by the end of the century under the RCP8.5 scenario. The annual delta changes of *P* and *PET* suggest that the latitudinal gradient noticed for both atmospheric fields in the historical period (Fig. 5a) tends to fade away, and thus *P* and *PET* become more homogeneous over the entire study area in the future periods.

The delta changes of *AET* range from significant increases of up to 30% in the northern subwatersheds to significant decreases of about 30% in the south for the period 2071-2100 under the RCP8.5 scenario (Fig. 6). The *AET* changes reflect a greater heterogeneity than the *P* and *PET* ones, and although the historical values follow a north-south distribution (Fig. 5a), the future *AET* changes do not compensate the latitudinal gradient. On the contrary, they exacerbate it, leading to a widening on the range of values of *AET*.

The delta changes of $Q$ are prevalently negative and statistically significant for all periods and RCPs, with changes below -40% in some of the southern subwatersheds for the period 2071-2100 under the RCP8.5 scenario (Fig. 6). In this respect, two main driving mechanisms for the annual $Q$ detriments were identified: 1) the future $P$ decreases in the southern mountains; and 2) the combined effect of the future $P$ reductions and the $AET$ increases in the north part of the basin. The former mechanism supposes that both the runoff generation and the evapotranspiration processes become limited by the less abundant precipitations under the future scenarios. The latter corresponds to those areas where there is still enough water availability for satisfying the higher atmospheric demand for water vapour (i.e. higher $PET$) in the future, and therefore represents a two-fold limiting factor for the runoff generation.

**4.3 Seasonal delta changes of *P, PET, AET* and *Q***

The mean seasonal values of *P*, *PET* and *AET* for the historical period are mapped in Fig. 5b, and Table 2 includes the mean seasonal values of *Q*. Seasonal *P* evidences a latitudinal gradient for all the seasons comparable to that observed in Fig. 5a for annual scale. The highest values are reached during autumn and are above 500 mm/season for various northern subwatersheds. The minimum values correspond to summer and are below 100 mm/season in the south. Seasonal *PET* is broadly below 100 mm/season in autumn and winter, being the latitudinal gradient almost inexistent. The *PET* values start to be noticeable in spring and achieve their maximum in summer with values above 500 mm/season in the southern subwatersheds. The maximum *AET* values were obtained for the spring months and are above 200 mm/season for the majority of the subwatesheds. The summer *AET* values are somewhat lower than the spring ones, and they are broadly below 200 mm/season in autumn and winter. Finally, the highest *Q* volumes take place in winter and are lower in autumn and spring, with minimums attained in summer (Table 2).

Fig. 6 also shows the seasonal delta changes of *P*, *PET*, *AET* and *Q* for the period 2071-2100 under the RCP8.5 scenario, being the changes associated to the rest of the future periods and scenarios shown together with their annual counterparts in the supplementary material (Fig. S3 to Fig. S7). The smallest decreases of seasonal *P* correspond to autumn, while they predominantly exceed 20% in spring and summer. Delta changes for winter, in turn, are mostly positive, and notably significant under the RCP4.5 scenario for 2071-2100 (Fig. S7) and the RCP8.5 scenario for 2041-2070 (Fig. S6). Little difference was found for the seasonal *PET* changes with respect to the annual changes, though the significant increments are slightly larger in autumn and winter, and less severe in spring and summer.

The significant *AET* increases detected for autumn and winter (Fig. 6) manifest that the evapotranspiration process is limited by the atmospheric demand of water vapour (i.e. *PET*) for the first half of the hydrologic year. During summer, however, the *AET* changes are negative and significant in the entire region, suggesting that the water availability constrains the evaporative fluxes. The spring *AET* changes lie between those extremes and show significant increases in the northern part of the basin and significant decreases over the south, with similar results for the rest of the future periods and scenarios (Fig. S3 to Fig. S7). Hence, the seasonal *AET* changes can explain the annual *AET* delta changes as follows: 1) the increases identified for the northern subwatersheds are due to the increments occurring in autumn, winter and spring, without a noteworthy effect of the summer diminutions on the annual differences; and 2) the projected detriments in the southernmost areas are promoted by the spring and summer decreases, whilst the autumn and winter increases do not cause a flip in the sign of the delta changes.

The seasonal delta changes of *Q* are mostly negative and more pronounced in spring and summer, reaching reductions above 40% in a great number of the subwatersheds for the RCP8.5 scenario during the period 2071-2100 (Fig. 6). There is a strong interplay between the seasonal

*P* and *AET* as the driving forces of the seasonal *Q* detriments. Thus, *P* constitutes the limiting factor for both the runoff generation and the evapotranspiration processes in summer, and over the southern part of the basin in spring. The combined effect of *P* decreases and *AET* increments is relevant in autumn, as well as for the northern subwatersheds in spring. The negative delta changes of *Q* for winter are related to the sharp and significant *AET* increments.

### 4.4 Delta changes of $Q_{snow}$

The application of Equation (1) allowed obtaining the $Q_{snow}$ monthly time series for each subwatershed of the study area that subsequently were aggregated at annual and seasonal time scales. Table 3 collects the mean annual and seasonal values of $Q_{snow}$ for the historical period, showing higher $Q_{snow}$ values for the northern subwatersheds as they are characterized by a greater elevation (see Fig. 1 and Table S1). Fig. 7 shows the delta changes of annual and seasonal $Q_{snow}$ for the period 2071-2100 under the RCP8.5 scenario. The remaining results for $Q_{snow}$ are shown in supplementary material (Fig. S8). The summer months were excluded from this analysis due to the absence of a snowmelt component for the streamflow values in this season.

The delta changes of $Q_{snow}$ are always negative and significant, mostly exceeding 80% in the long-term future period (Fig. 7), and being more pronounced in autumn and spring. $Q_{snow}$ constitutes the hydrological variable for which the impact of climate change is more evident, evincing a generalized tendency for the headwaters of the Duero River Basin to be much less snow dominated as a consequence of climate change.

### 4.5 Future changes of $i_Q$, $i_E$ and $Q_{snow}$ ratio

Fig. 8a shows the values of the signature measures $i_Q$ and $i_E$ for the historical period, and their future changes are collected in Fig. 8b for the period 2071-2100 under the RCP8.5 scenario (Fig. S9 in the supplementary material depicts the results of $i_Q$ for the rest of the periods and RCP scenarios). The highest values of $i_Q$ and $i_E$ occur for the northern and for the southern subwatersheds, respectively, being the sum of both ratios always close to 1. This complementary assumption is also applicable to the $i_Q$ and $i_E$ changes, being their sum close to 0. The future changes of $i_Q$ and $i_E$ suggest that the northern subwatersheds tend towards the evaporative regime range, and the evaporative response becomes stronger in the south.

The $Q_{snow}$ ratio values for the historical period are presented in Fig. 9a at both annual and seasonal scales, excluding summer. Most of the annual $Q_{snow}$ ratio values are above 0.3 for the northern basins, being weaker in the southern mountains, with values that mainly range from 0.1 to 0.3. The seasonal distribution reveals that the highest $Q_{snow}$ ratios are concentrated in the winter months and exceeds 0.4 in many northern subwatersheds. Autumn and spring show similar results with values always below 0.3. The future changes of the $Q_{snow}$ ratio at annual and seasonal time scales are depicted in Fig. 9b for the period 2071-2100 under the RCP8.5 scenario (the rest of the changes are shown in Fig. S10 in the supplementary material), and manifest a clear predominance of negative values broadly below -0.1.

**4.6 Future changes of the centroid position**

Table 4 collects the coordinate pairs ($C_X$, $C_Y$) of the annual and seasonal centroids for the daily hydrograph for the average hydrologic year in the historical period. The annual $C_X$ values present a mean value of 150.4 days and a difference of 41.9 days between the maximum and minimum values (i.e. dispersion). The lowest annual $C_X$ values were generally obtained for northern subwatersheds (e.g. R-2027, R-2028 and GS-3150), being the highest ones mainly located in the south (e.g. R-2037, R-2043 and GS-3057). The seasonal $C_X$ values are less dispersed, with a maximum difference of 13.1 days in autumn and not exceeding 10 days in winter and spring. The spatial distribution of $C_X$ for autumn and spring is comparable to that obtained for the entire year, while spring exhibits an opposite pattern with minimums attained in the south. On the other hand, a strong correlation ($r > 0.99$) was found between the $C_Y$ measures and the average $Q$ values for the historical period (Table 2) in all cases, thus implying that the highest $C_Y$ values take place in winter and are lower in autumn and spring.

The projected changes of the centroid position are shown in Fig. 10. The annual $C_X$ changes (Fig. 10a) are predominantly negative and more pronounced for the RCP8.5 scenario, being the differences mostly below 10 days. A similar behaviour is also noticeable, but to a lesser degree, in winter and spring. The autumn $C_X$ changes are, in turn, prevailingly positive and present increases of up to 5 days. This represents an important feature of the future behaviour of the autumn streamflow that is not well-captured in the annual $C_X$ changes.

The $C_Y$ changes (Fig. 10b) are negative and broadly below 20% both at annual and seasonal time scales. However, the spring $C_Y$ decreases are more noticeable and usually exceed 20%, reaching decreases above 40% for RCP8.5. The correlation between the $C_Y$ changes and the delta changes of $Q$ (Table S2 in the supplementary material) is characterized by $r$ values statistically significant at a 95% confidence level, with a varying degree of correlation depending on the time scale and RCP considered. The $C_Y$ and the $Q$ changes practically show a 1:1 relationship for winter and spring and both RCPs. Conversely, the correlation is less marked for autumn and for the complete hydrologic year, with $r$ values below 0.9 and slightly above 0.9 at annual scale for the RCP8.5 scenario. Therefore, as the delta changes of $Q$, the projected changes of $C_Y$ pinpoint a generalized decrease of the streamflow for the study area, being in some cases interchangeable measures (i.e. in winter and spring).

## 5. Discussion

### 5.1 Projected annual hydrologic changes

A similar spatial pattern for the annual *P* changes over the Duero Basin (Fig. 6) was detected using WRF simulations over Spain in Argüeso et al. (2012), concluding that the changes tend to be larger in the southern half of the domain, particularly over the mountainous areas. Likewise, the annual *P* changes are in agreement with those obtained in Fonseca and Santos (2019) for the Tâmega River using also a Euro-CORDEX ensemble. The projected annual *PET* changes are coherent with the findings of Moratiel et al. (2011) for the Duero Basin, where increases between 5% and 11% are expected for the annual *PET* by the end of the first half of the 21$^{st}$ century.

The annual delta changes of *Q* and *AET* (Fig. 6) corroborate the sign and the magnitude of the projected changes of annual *Q* and *AET* for the Tâmega Basin in Fonseca and Santos (2019), as well as they further extend the conclusions reached there given the greater number of subwatersheds considered in our study and the higher number of members included in the Euro-CORDEX ensemble. Nonetheless, this study is focused in the Spanish part of the basin, and even though it can be considered as representative of the entire area, future research will encompass the totality of the basin in order to overcome this limitation.

On the other hand, and similarly to this study, negative $i_Q$ changes were found in Mendoza et al. (2015, 2016) in the Colorado River Basin. The steady-state assumption reflected in Eq. (3) is one well-known and widespread approach taken for the quantitative analysis of the water balance equation (e.g. Rasmussen et al., 2014; Xu et al., 2014; Liang et al., 2015; Mendoza et al., 2015, 2016; Hasan et al., 2018; Li et al., 2018). However, this assumption is rarely checked and can lead to a long-term imbalance when the storage component is not considered even for long periods (i.e. 10 to 30 years), particularly in arid and semi-arid regions (Han et al., 2020). In order to avoid feasible inaccuracies in the application of this approach, the steady-state assumption was tested for all the studied subwatersheds, remaining the sum of $i_Q$ and $i_E$ close to 1 for the historical period and for all the future scenarios.

**5.2 Projected seasonal hydrologic changes**

The statistically significant positive delta changes corresponding to winter *P* (Fig. 6) point to an important feature for the future precipitations in the central subwatersheds of the Duero Basin. Similarly, other studies using WRF projections concluded that, contrarily to the rest of the seasons, winter precipitation is projected to increase over certain areas of the Iberian Peninsula due to climate change, remarkably over the Northern Plateau, but the increases are generally non-significant (Argüeso et al., 2012; García-Valdecasas et al., 2020a). In the same vein, Soares et al. (2017), using WRF and Euro-CORDEX ensembles, found both significant and non-significant increases of winter precipitation for some areas in the north of Portugal including the Portuguese part of the Duero River Basin.

The findings for the seasonal *AET* changes (Fig. 6) partially agree with the results reported in García-Valdecasas et al. (2020a), where comparable changes were found during winter and summer over the study area. However, the WRF simulations carried out there diverged from our results for the rest of the hydrologic year: in autumn, the WRF projections led to significant decreases for almost all the simulations, and in spring, the partitioning between significant increases in the northern headwaters and significant decreases in the south were not captured. It is expected that the VIC implementation of this work reproduces more realistically the water balance and the future hydrologic changes of the Duero headwaters since it was built upon the calibration exercise developed in Yeste et al. (2020). Although the model was calibrated only using streamflow observations, its performance was also evaluated against two *AET* products, producing good adjustments as well and improving the benchmark performance in all cases. Other feasible explanation can be related to the high number of models included in the Euro-CORDEX ensemble in comparison to the two GCMs that drove the WRF simulations in García-Valdecasas et al. (2020a). Lastly, the choice of the climatological year (i.e. from December to November) in García-Valdecasas et al. (2020a) could also conduct to some differences when compared to the results for the hydrologic year (i.e. from October to September) presented in this work.

The projected seasonal changes for $Q$ (Fig. 6) were similar to those obtained in Fonseca and Santos (2019) for the Tâmega Basin, with downward trends for all the seasons except for winter, where a slight increase was projected. The strongest diminutions were projected for summer, where water scarcity and the increasing frequency of droughts may pose a serious threat in the future in agreement with García-Valdecasas Ojeda et al. (2021). Although most of the summer $Q$ changes were characterized by marked and significant decreases in our study, it is important to note the existence of a few significant increases (Fig. 9). This responds to an atypical behaviour and is presumably driven by two factors: firstly, the very nature of the low $Q$ values in summer supposes that a higher streamflow, though remaining in the low range, produces a markedly positive delta change; and secondly, the averaging of all the $Q$ time series when the mean of the ensemble was calculated could introduce small biases that finally led to a positive delta change in rare cases.

The results for $Q_{snow}$ (Fig. 7) and $Q_{snow}$ ratio (Fig. 9) resemble the relative contribution of the snowmelt component to the generated runoff in the Ganges basin applying an identical method for estimating $Q_{snow}$ (Siderius et al., 2013), which was expected to change as climate warms. Similarly, Ceballos-Barbancho et al. (2008) and Morán-Tejeda et al. (2010, 2011) pointed to a change of the snow regime in the Duero River Basin during the last half of the 20$^{th}$ century, with important implications for water management that already led to the adoption of different management practices in other parts of Spain (López-Moreno et al., 2004). The marked reductions observed for winter and spring streamflow were likely caused by the decrease of winter precipitation and the increase of winter and spring temperatures. The latter implies a decrease of snow accumulation in winter and an earlier snowmelt presence during spring, therefore affecting the amount and timing of the streamflow (Morán-Tejeda et al., 2010, 2011). This downward tendency driven by a warmer climate has being previously identified for the mountainous areas in Spain (López-Moreno et al., 2009; Morán-Tejeda et al., 2017; Collados-Lara et al., 2019) and other parts of the world (e.g. Bhatti et al., 2016; Majone et al., 2016; Coppola et al., 2018; Ishida et al., 2018, 2019; Liu et al., 2018), thus suggesting a critical role of the snowmelt component for the future management of mountain water resources (Viviroli et al., 2011; Mankin et al., 2015).

**5.3 Annual and seasonal changes of the hydrograph centroid**

The projected annual changes of $C_X$ (Fig. 10a) suggest that a time shift in the hydrologic year towards earlier streamflow volumes takes place for the future scenarios. The "center time" of runoff is considered a measure of the streamflow seasonality, and is usually calculated for the average hydrologic year as a single metric for the entire hydrograph (Stewart et al., 2005; Mendoza et al., 2015; Mendoza et al., 2016). Mendoza et al. (2015) suggested that the negative sign of the projected annual changes of $C_X$ are linked to a lesser presence of snow under climate change conditions. This is consistent with the findings of this work for the studied subwatersheds, where $Q_{snow}$ is expected to suffer the greatest burden of the impacts of climate change.

It is important to note that with the only use of the annual $C_X$ position as a signature measure there are other important characteristics of the average daily hydrograph that remain hidden and not completely depicted. This limitation has been overcome by calculating the seasonal centroid position and its future changes (Fig. 10), revealing additional information about the seasonal timing of the streamflow that is expected to have a large impact on the future water management strategies for the basin.

Finally, the annual and seasonal $C_Y$ changes (Fig. 10b) constitute another valuable metric that can be related to the delta changes of $Q$ both at annual and seasonal scales (Table S2). The degree of correlation between them responds to the question of to which extent the changes of the shape of the hydrograph (i.e. $C_Y$ changes) are related to the changes of the mean values (i.e. delta changes of $Q$). Thus, the 1:1 relationship observed for winter and spring indicates that the changes of shape are mainly driven by the delta changes of $Q$. This is also supported by the closeness to 0 of the $C_X$ changes for these seasons (Fig. 10a), being the shape of the hydrograph directly related to the vertical shifting of the centroid. The annual and autumn changes of the centroid position show greater complexity as the $C_X$ changes become a contributing factor to the changes of shape. In these cases the $C_Y$ changes and the delta changes of $Q$ are less correlated but still manifest a linear relationship with statistically significant $r$ values. This approach generalizes the common usage of the "center time" of runoff as a measure of seasonality, and further research will explore the implications of the changes of the $C_X$ and $C_Y$ and their relation to the corresponding delta changes for the rest of atmospheric and land variables involved in the hydrology of the headwaters of the Duero River Basin.

## 6. Conclusions

The multi-model ensemble approach has shown to be an effective tool for the analysis of the impacts of climate change in the headwater areas of the Duero River Basin both at annual and seasonal time scales. The simulations carried out with the VIC model driven by a large number of Euro-CORDEX RCM+GCM combinations and two RCPs has permitted a posterior analysis applying the delta change method and estimating various signature measures for the different land and atmospheric variables enmeshed in the modelling exercise. The former evaluated the future changes of the mean values, and the latter addressed other important hydrologic features including the relative contribution of runoff and actual evapotranspiration to the overall water balance, the snowmelt contribution to the total streamflow and the centroid position for the daily hydrograph of the average hydrologic year. The main findings of this work are as follows:

- The annual streamflow decreases were driven by two different mechanisms depending on the mountainous system considered. The precipitation decreases in the south part of the basin imposed a limit to the runoff and evapotranspiration processes. The streamflow reductions for the northern mountains were the outcome of a combined effect of the precipitation decreases and evapotranspiration increases in the future scenarios. The future changes of the runoff and the evapotranspiration ratios revealed a tendency towards an evaporative regime for the northern subwatersheds, while the evaporative response strengthened in the south. The sum of both ratios remained close to 1 for all the studied cases, thus confirming the steady-state assumption usually non-tested in many previous studies.

- The precipitation and evapotranspiration changes evinced a strong intra-annual variability, and were directly related to the seasonal streamflow detriments: the precipitation decreases constituted the limiting factor for the runoff and evapotranspiration processes in summer for all the studied subwatersheds, and over the southern part of the basin in spring; the compound effect of the precipitation reductions and the evapotranspiration increments was noticeable in autumn for the entire basin, and over the north in spring; lastly, the winter streamflow changes were mostly negative and non-significant as a consequence of the non-significant changes projected for the precipitation in this season.

- The snowmelt contribution to the total generated runoff was the hydrologic variable most affected by the climate warming over the study area. The projected changes indicated a downward tendency towards the practically non-existence of snow dominated hydrologic regimes for the headwaters of the Duero River Basin. This behaviour exacerbates the previous findings for the mountainous areas in Spain during the last half of the 20$^{th}$ century, and suggests a major role of this component for the future water management practices.

- The projected changes of the centroid position were estimated for the average daily hydrologic year at annual and seasonal scales, and accounted for the variations of the streamflow seasonality (i.e. horizontal shifts) and the streamflow volumes (i.e. vertical shifts). Particularly, the vertical shifts showed a strong degree of correlation to the corresponding delta changes of the streamflow, being interchangeable measures in winter and spring. This approach generalized the widespread use of the "center time" of runoff as a signpost of seasonality as many other key features were well-captured and fully explained, and can be further applied for the rest of atmospheric and land variables involved in the modelling exercise.

**Acknowledgements**

All the simulations were conducted in the ALHAMBRA cluster (http://alhambra.ugr.es) of the University of Granada. This work was funded by the Spanish Ministry of Economy and Competitiveness projects CGL2013-48539-R and CGL2017-89836-R, with additional support from the European Community Funds (FEDER) and by FEDER/Junta de Andalucía-Consejería de Economía y Conocimiento/B-RNM-336-UGR18 project. The first author was supported by the Ministry of Education, Culture and Sport of Spain (FPU grant FPU17/02098). We thank three anonymous referees whose comments and indications improved the paper significantly.

**List of figures**

**Figure 1.** Duero River Basin and the 24 studied subwatersheds. The prefix "R-" denotes "Reservoir" and the prefix "GS-" denotes "Gauging Station".

**Figure 2.** X and Y coordinates of the centroid for the daily hydrograph ($C_X$, $C_Y$) at annual and seasonal scales: a) annual centroid, b) fall centroid, c) winter centroid and d) spring centroid.

**Figure 3.** CDFs of the VIC model performance for the period Oct 2000 – Sep 2011 corresponding to the combination historical+RCP8.5: a) *NSE* for the streamflow simulations against streamflow observations; b) *NSE* for the *AET* simulations against GLEAM; c) $i_Q$ bias with respect to the ratio of SPREAD precipitation to streamflow observations. Blue lines represent the ensemble simulation, and orange lines correspond to the calibration results from Yeste et al. (2020).

**Figure 4.** CDFs of the VIC model performance for the annual and seasonal values of $C_X$ (a to d) and $C_Y$ (e to h) corresponding to the period Oct 2000 – Sep 2011 and the combination historical+RCP8.5. $C_X$ biases are calculated as the difference between simulated and observed values. $C_Y$ biases represent fractional changes calculated as [(simulations−observations)/observations × 100]. Blue lines represent the ensemble simulation, and orange lines correspond to the calibration results from Yeste et al. (2020).

**Figure 5.** Basin-averaged annual and seasonal values of *P*, *AET* and *PET* for the historical period.

**Figure 6.** Delta changes of annual and seasonal *P*, *Q*, *AET* and *PET* for the period 2071-2100 under the RCP8.5 scenario in the studied subwatersheds. Significant changes at the 95% confidence level have been marked with solid borders.

**Figure 7.** Delta changes (excluding summer) of annual and seasonal $Q_{snow}$ for the period 2071-2100 under the RCP8.5 scenario in the studied subwatersheds. Significant changes at a 95% confidence level have been marked with solid borders.

**Figure 8.** (a) Values of $i_Q$ and $i_E$ for the historical period in the studied subwatersheds. (b) Projected changes of $i_Q$ and $i_E$ calculated as the difference between future and historical values for the period 2071-2100 under the RCP8.5 scenario.

**Figure 9.** (a) Annual and seasonal values (excluding summer) of $Q_{snow}$ ratio for the historical period in the studied subwatersheds. (b) Projected changes of annual and seasonal $Q_{snow}$ ratio (excluding summer) calculated as the difference between future and historical values for the period 2071-2100 under the RCP8.5 scenario.

**Figure 10.** Projected changes of the centroid position at annual and seasonal time scales for the different RCPs: a) changes of $C_X$ calculated as the difference between future and historical values, and b) fractional changes [(future−historical)/historical × 100] of $C_Y$. Boxplots in a) and b) represent the 24 studied subwatersheds.

**Table 1.** Ensemble of combinations RCM+GCM chosen from the Euro-CORDEX database

| RCM | GCM | RCM | GCM |
|---|---|---|---|
| RCA4 | CNRM-CERFACS-CNRM-CM5 | HIRHAM5 | ICHEC-EC-EARTH, r3i1p1 |
| RCA4 | ICHEC-EC-EARTH, r12i1p1 | HIRHAM5 | NCC-NorESM1-M |
| RCA4 | IPSL-IPSL-CM5A-MR | RACMO22E | CNRM-CERFACS-CNRM-CM5 |
| RCA4 | MOHC-HadGEM2-ES | RACMO22E | ICHEC-EC-EARTH, r12i1p1 |
| RCA4 | MPI-M-MPI-ESM-LR | RACMO22E | ICHEC-EC-EARTH, r1i1p1 |
| CCLM4-8-17 | CNRM-CERFACS-CNRM-CM5 | REMO2009 | MOHC-HadGEM2-ES |
| CCLM4-8-17 | ICHEC-EC-EARTH | REMO2009 | MPI-M-MPI-ESM-LR, r1i1p1 |
| CCLM4-8-17 | MOHC-HadGEM2-ES | REMO2009 | MPI-M-MPI-ESM-LR, r2i1p1 |
| CCLM4-8-17 | MPI-M-MPI-ESM-LR | WRF331F | IPSL-IPSL-CM5A-MR |

**Table 2.** Basin-averaged annual and seasonal values of *Q* for the historical period.

| Code | Annual (hm$^3$/year) | Autumn (hm$^3$/season) | Winter (hm$^3$/season) | Spring (hm$^3$/season) | Summer (hm$^3$/season) |
|---|---|---|---|---|---|
| R-2001 | 312.22 | 65.43 | 121.62 | 101.05 | 24.12 |
| R-2011 | 93.70 | 21.41 | 40.31 | 29.22 | 2.76 |
| R-2013 | 150.37 | 42.42 | 57.24 | 41.88 | 8.83 |
| R-2014 | 229.93 | 60.93 | 84.57 | 69.71 | 14.71 |
| R-2026 | 445.30 | 110.75 | 158.98 | 138.26 | 37.31 |
| R-2027 | 23.32 | 7.44 | 10.90 | 4.36 | 0.62 |
| R-2028 | 71.66 | 25.41 | 24.70 | 17.98 | 3.57 |
| R-2032 | 658.43 | 184.00 | 266.30 | 181.26 | 26.86 |
| R-2036 | 46.80 | 11.21 | 17.89 | 15.00 | 2.70 |
| R-2037 | 99.36 | 22.37 | 40.16 | 25.90 | 10.93 |
| R-2038 | 773.35 | 237.64 | 337.55 | 178.98 | 19.18 |
| R-2039 | 299.90 | 88.51 | 151.97 | 55.34 | 4.07 |
| R-2042 | 122.79 | 22.17 | 64.59 | 32.49 | 3.55 |
| R-2043 | 87.76 | 16.84 | 33.61 | 31.56 | 5.75 |
| GS-3005 | 113.82 | 19.94 | 55.21 | 30.53 | 8.13 |
| GS-3016 | 81.50 | 17.42 | 36.68 | 22.49 | 4.90 |
| GS-3041 | 16.69 | 2.98 | 9.56 | 3.79 | 0.35 |
| GS-3049 | 19.12 | 2.89 | 7.64 | 7.18 | 1.41 |
| GS-3051 | 18.38 | 3.66 | 8.43 | 5.72 | 0.58 |
| GS-3057 | 47.05 | 8.83 | 21.39 | 15.20 | 1.63 |
| GS-3089 | 182.53 | 46.51 | 79.67 | 46.39 | 9.96 |
| GS-3104 | 147.60 | 36.24 | 64.15 | 36.22 | 11.00 |
| GS-3150 | 191.03 | 57.94 | 73.14 | 53.74 | 6.20 |
| GS-3818 | 270.69 | 52.50 | 131.59 | 74.83 | 11.77 |

**Table 3.** Basin-averaged annual and seasonal values (excluding summer) of $Q_{snow}$ for the historical period.

| Code | Annual (hm$^3$/year) | Autumn (hm$^3$/season) | Winter (hm$^3$/season) | Spring (hm$^3$/season) |
|---|---|---|---|---|
| R-2001 | 69.44 | 11.00 | 44.61 | 13.82 |
| R-2011 | 22.25 | 4.24 | 14.44 | 3.56 |
| R-2013 | 35.75 | 7.88 | 22.96 | 4.91 |
| R-2014 | 72.94 | 15.34 | 40.34 | 17.26 |
| R-2026 | 141.16 | 27.52 | 75.29 | 38.28 |
| R-2027 | 2.81 | 0.67 | 2.04 | 0.09 |
| R-2028 | 23.06 | 6.37 | 11.70 | 4.99 |
| R-2032 | 195.16 | 40.66 | 114.07 | 40.42 |
| R-2036 | 5.42 | 1.10 | 3.78 | 0.54 |
| R-2037 | 10.83 | 1.93 | 7.86 | 1.03 |
| R-2038 | 122.19 | 27.18 | 82.45 | 12.56 |
| R-2039 | 6.80 | 1.67 | 5.07 | 0.05 |
| R-2042 | 19.50 | 2.73 | 15.42 | 1.35 |
| R-2043 | 27.99 | 4.71 | 16.09 | 7.19 |
| GS-3005 | 9.77 | 1.30 | 8.13 | 0.34 |
| GS-3016 | 11.47 | 2.18 | 8.27 | 1.02 |
| GS-3041 | 1.29 | 0.15 | 1.12 | 0.02 |
| GS-3049 | 0.87 | 0.14 | 0.69 | 0.04 |
| GS-3051 | 5.02 | 0.85 | 3.37 | 0.80 |
| GS-3057 | 11.24 | 1.50 | 7.17 | 2.58 |
| GS-3089 | 39.49 | 7.80 | 26.75 | 4.94 |
| GS-3104 | 22.28 | 4.02 | 16.62 | 1.64 |
| GS-3150 | 60.15 | 13.67 | 32.81 | 13.63 |
| GS-3818 | 6.87 | 0.87 | 5.75 | 0.25 |

**Table 4.** Centroid position at annual and seasonal time scales in the daily hydrograph for the average hydrologic year corresponding to the historical period. $C_X$ is expressed in days since Oct 1 and $C_Y$ in hm$^3$/day.

| Code | Annual | | Autumn | | Winter | | Spring | |
|---|---|---|---|---|---|---|---|---|
| | $C_X$ | $C_Y$ | $C_X$ | $C_Y$ | $C_X$ | $C_Y$ | $C_X$ | $C_Y$ |
| R-2001 | 162.3 | 0.560 | 57.2 | 0.422 | 139.3 | 0.684 | 221.7 | 0.597 |
| R-2011 | 150.6 | 0.191 | 62.9 | 0.164 | 136.2 | 0.224 | 219.4 | 0.181 |
| R-2013 | 148.1 | 0.267 | 56.4 | 0.266 | 136.7 | 0.318 | 222.5 | 0.241 |
| R-2014 | 152.7 | 0.405 | 56.4 | 0.381 | 137.5 | 0.469 | 221.6 | 0.411 |
| R-2026 | 158.5 | 0.763 | 56.3 | 0.689 | 138.7 | 0.884 | 220.9 | 0.827 |
| R-2027 | 131.2 | 0.051 | 61.4 | 0.054 | 133.5 | 0.062 | 219.4 | 0.027 |
| R-2028 | 137.0 | 0.128 | 52.9 | 0.149 | 135.7 | 0.138 | 221.3 | 0.107 |
| R-2032 | 144.6 | 1.254 | 58.0 | 1.198 | 136.8 | 1.478 | 218.7 | 1.137 |
| R-2036 | 155.2 | 0.086 | 54.9 | 0.069 | 139.4 | 0.100 | 218.6 | 0.095 |
| R-2037 | 160.3 | 0.169 | 55.5 | 0.137 | 137.2 | 0.224 | 219.4 | 0.158 |
| R-2038 | 134.3 | 1.578 | 56.6 | 1.502 | 134.8 | 1.900 | 215.8 | 1.207 |
| R-2039 | 129.6 | 0.738 | 64.4 | 0.716 | 131.9 | 0.890 | 213.9 | 0.395 |
| R-2042 | 150.1 | 0.278 | 62.8 | 0.170 | 137.5 | 0.361 | 216.3 | 0.215 |
| R-2043 | 164.8 | 0.164 | 60.9 | 0.120 | 138.8 | 0.187 | 222.3 | 0.184 |
| GS-3005 | 157.8 | 0.233 | 58.9 | 0.136 | 137.0 | 0.314 | 217.1 | 0.199 |
| GS-3016 | 154.1 | 0.157 | 61.0 | 0.124 | 136.5 | 0.204 | 221.1 | 0.133 |
| GS-3041 | 144.3 | 0.043 | 66.0 | 0.027 | 134.0 | 0.056 | 214.6 | 0.027 |
| GS-3049 | 171.5 | 0.037 | 60.9 | 0.021 | 139.9 | 0.043 | 222.9 | 0.041 |
| GS-3051 | 153.6 | 0.038 | 64.4 | 0.029 | 137.3 | 0.047 | 219.4 | 0.035 |
| GS-3057 | 156.1 | 0.097 | 59.0 | 0.059 | 140.3 | 0.120 | 218.1 | 0.097 |
| GS-3089 | 147.5 | 0.352 | 59.9 | 0.320 | 135.9 | 0.445 | 219.2 | 0.285 |
| GS-3104 | 150.6 | 0.272 | 56.2 | 0.225 | 136.7 | 0.358 | 218.0 | 0.229 |
| GS-3150 | 141.6 | 0.367 | 58.0 | 0.379 | 136.2 | 0.407 | 219.6 | 0.337 |
| GS-3818 | 152.8 | 0.574 | 64.9 | 0.435 | 136.3 | 0.736 | 219.2 | 0.459 |

Figure 1

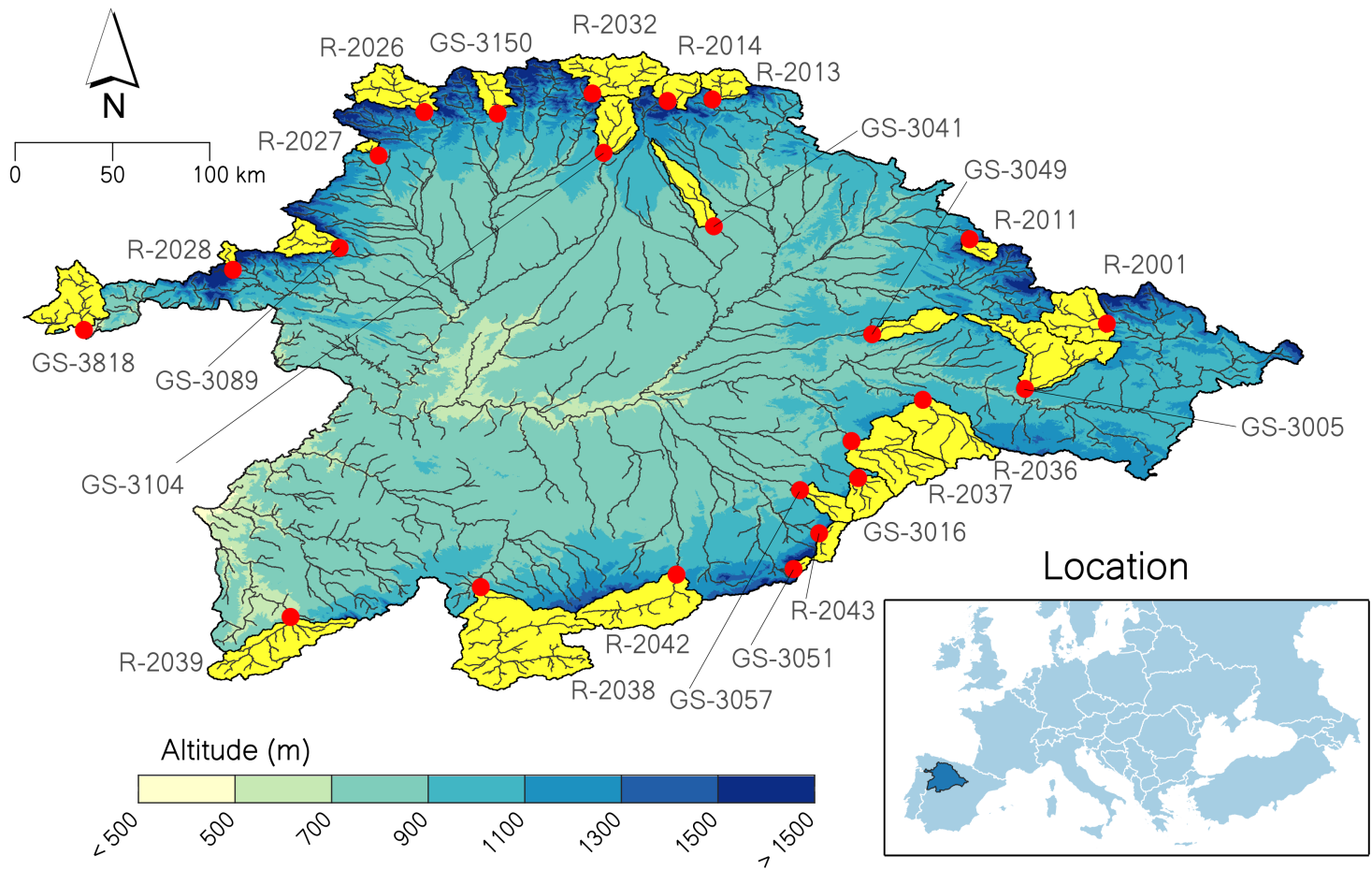

Figure 2

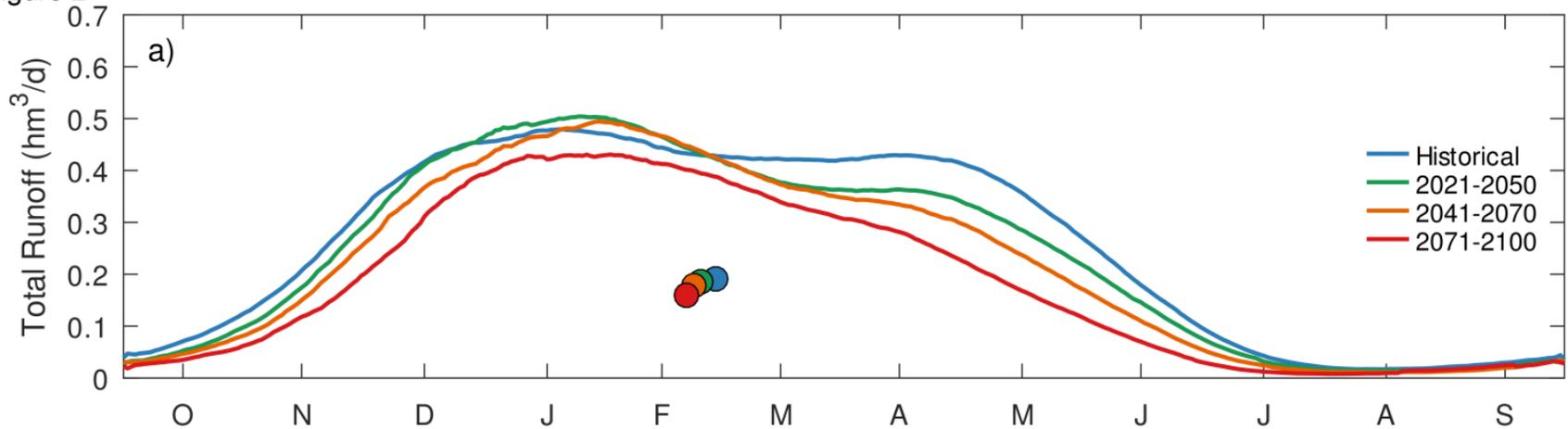
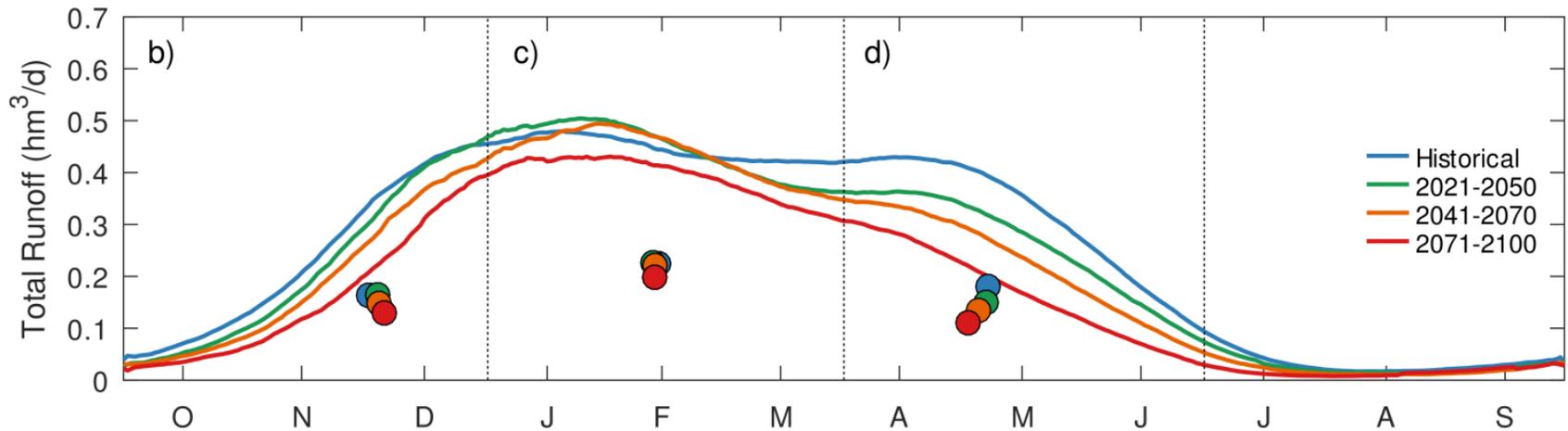

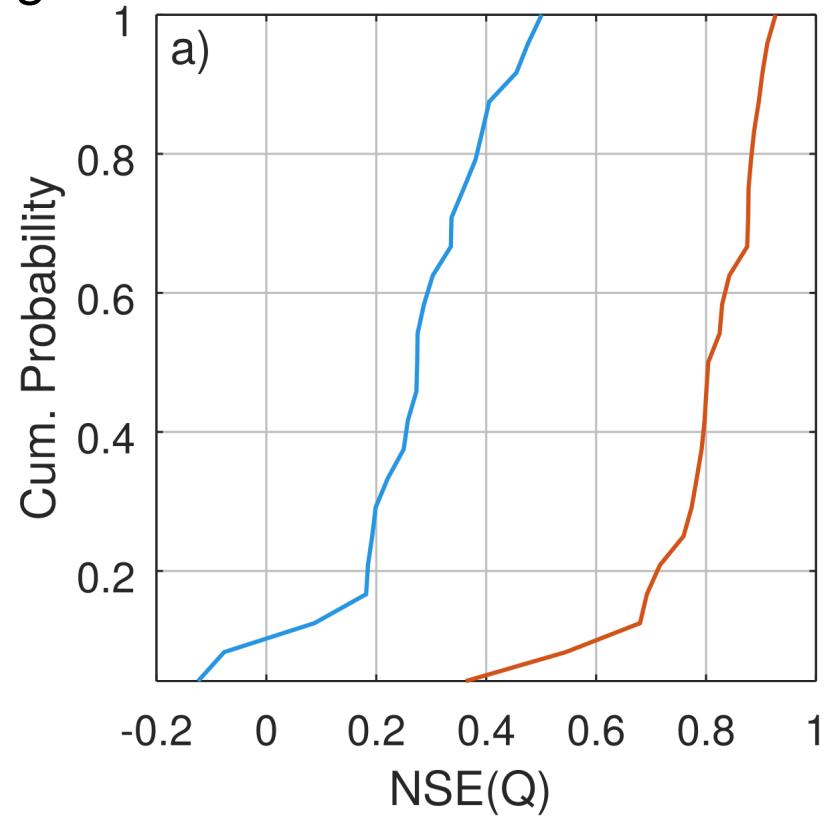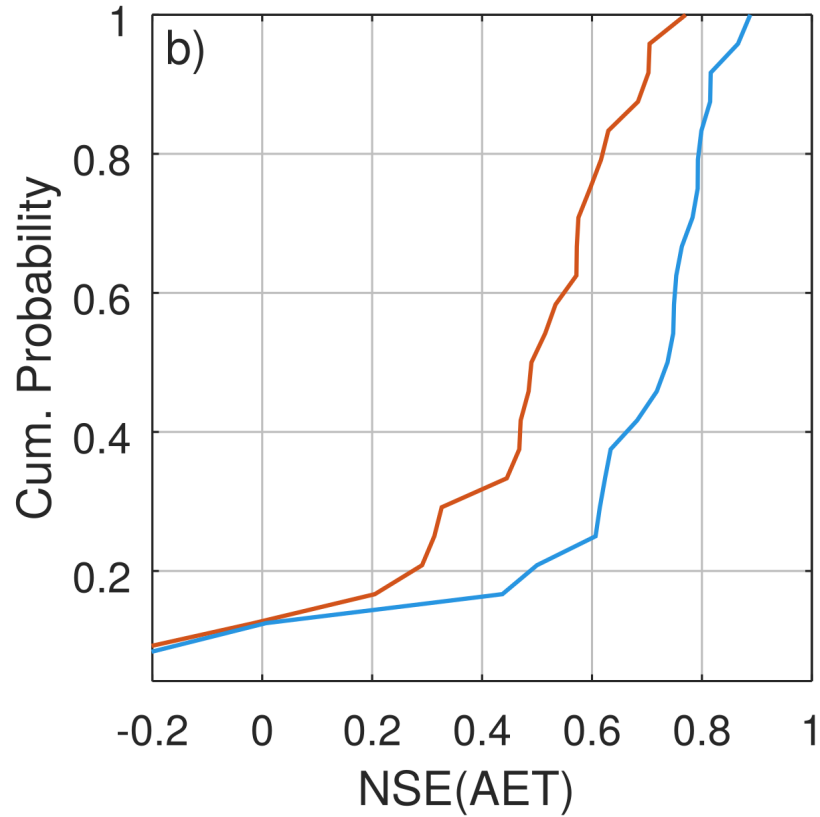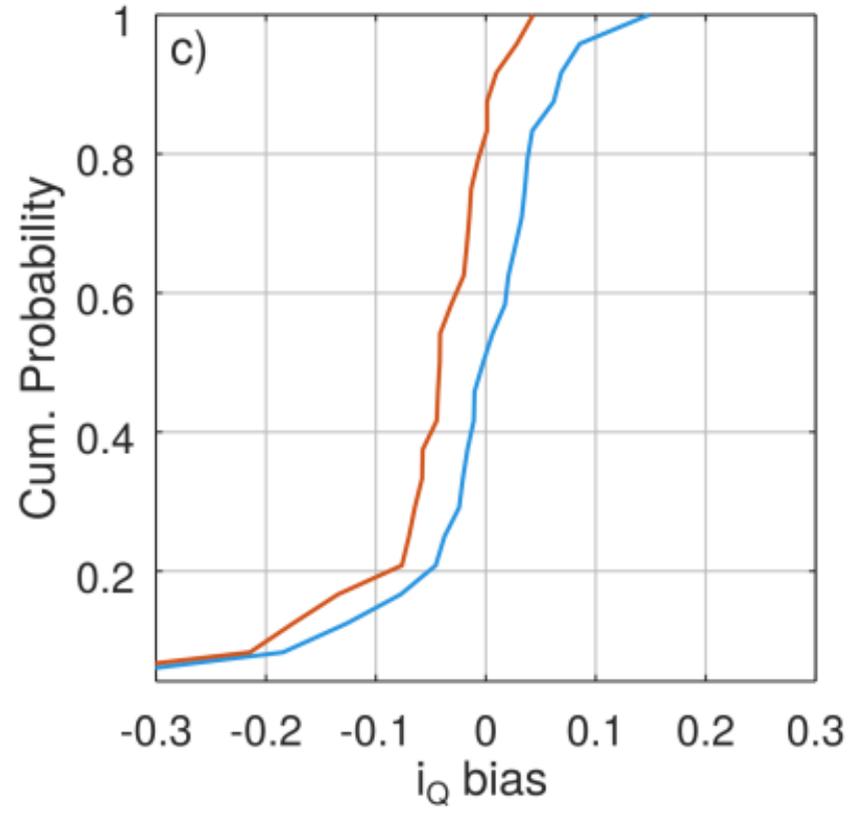

Figure 3

Figure 4

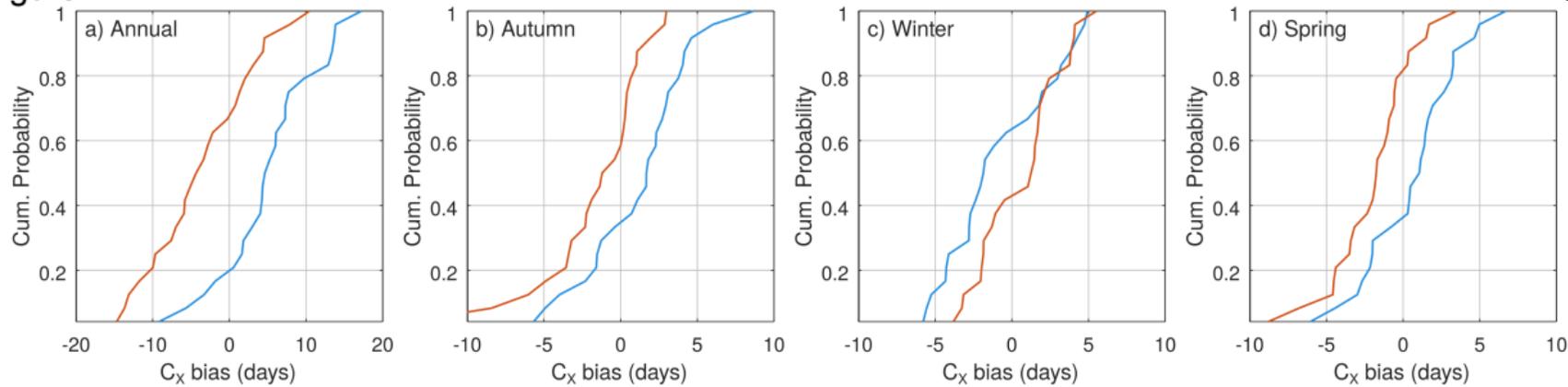
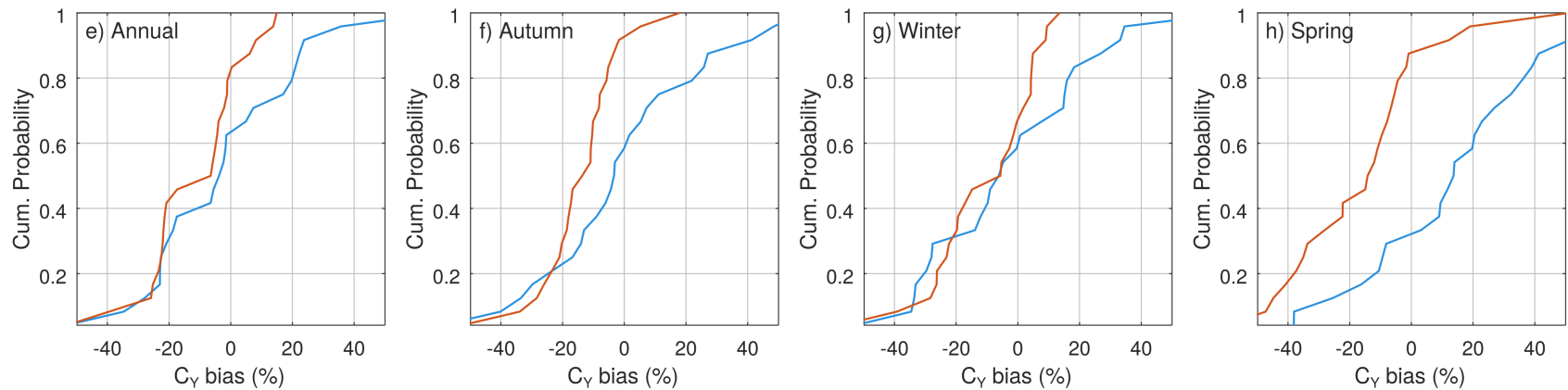

Figure 5

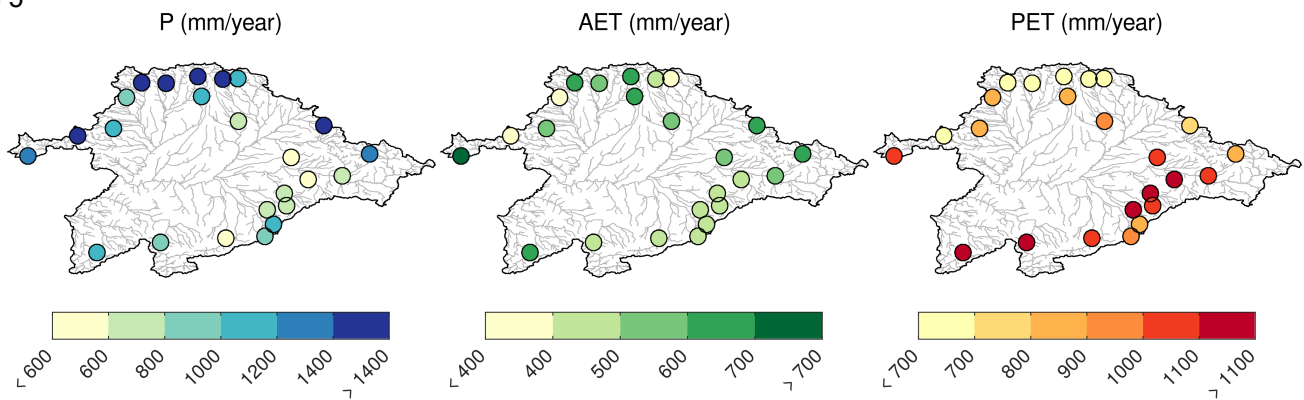
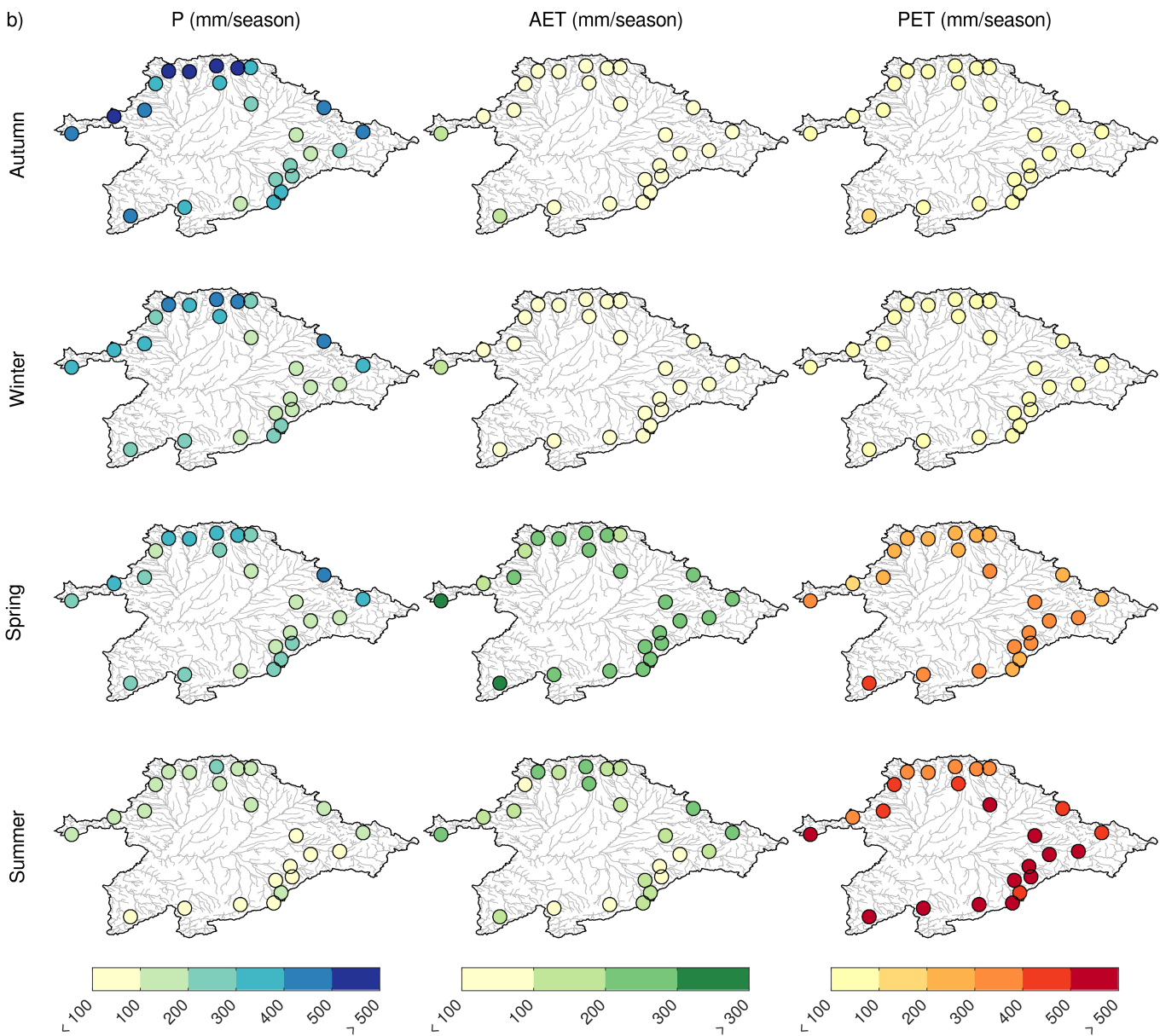

Figure 6

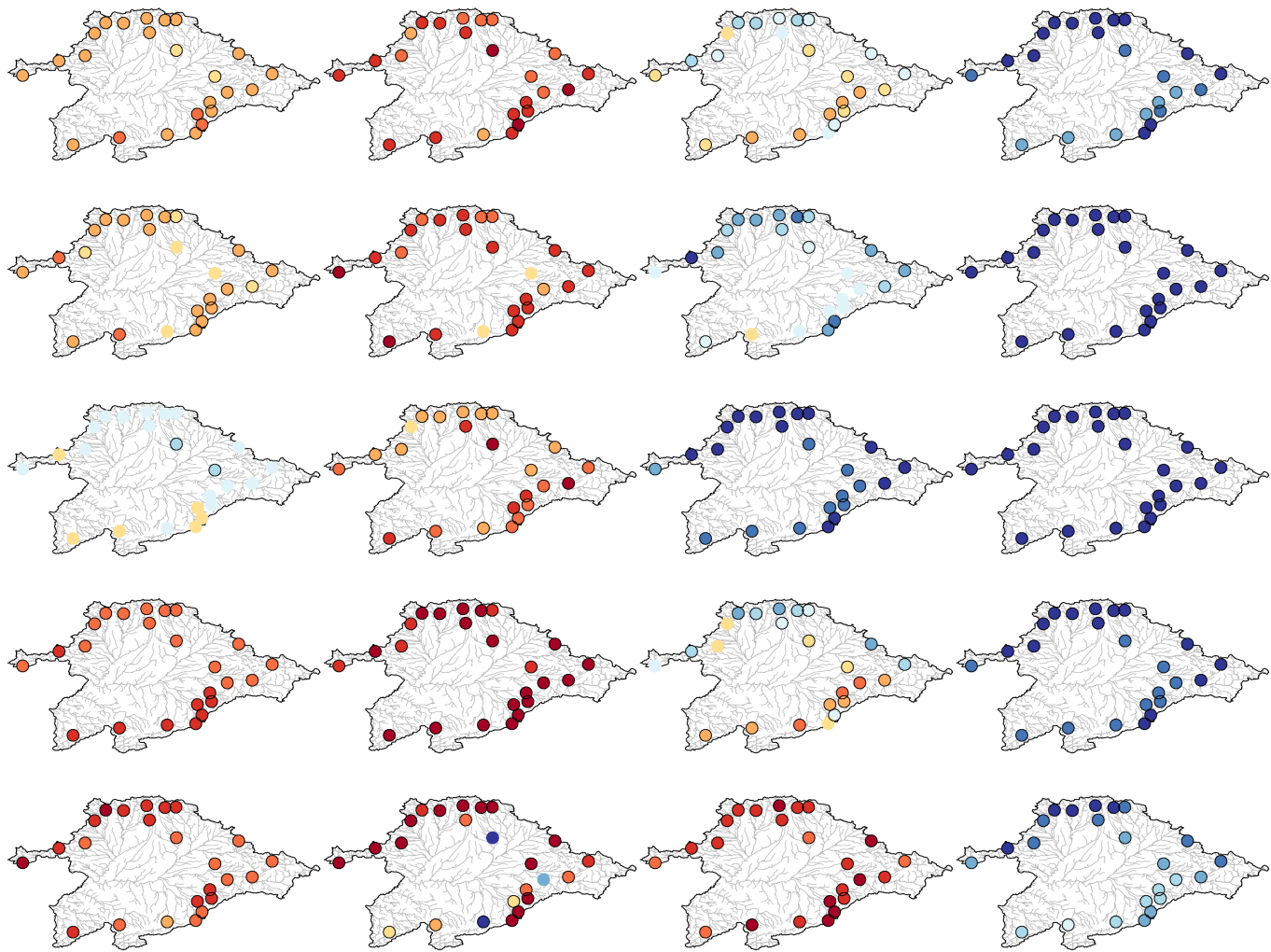

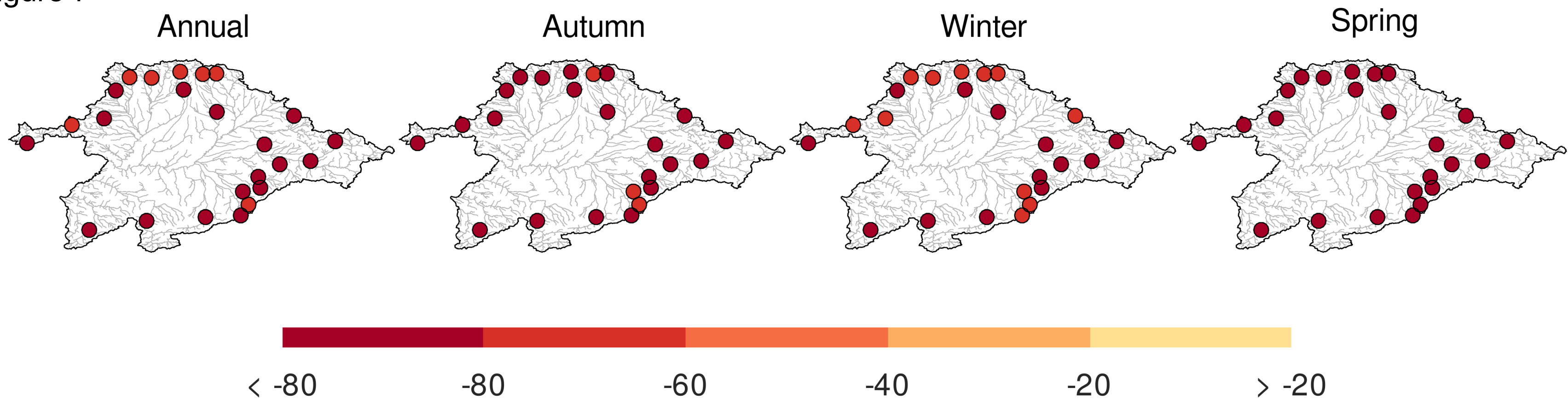

Figure 8

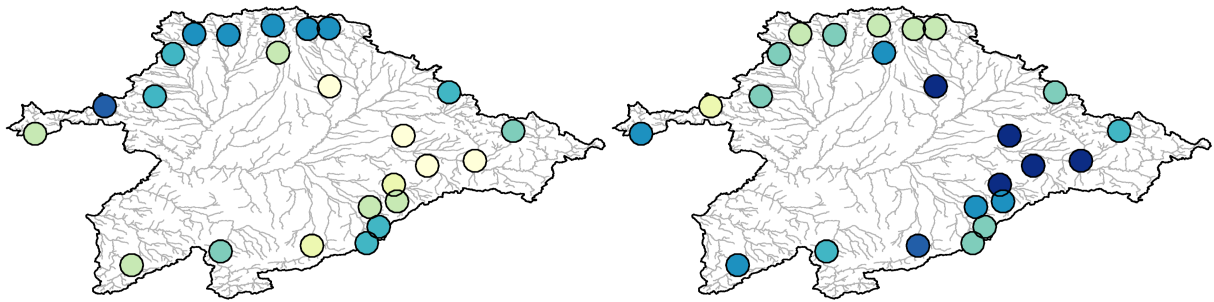
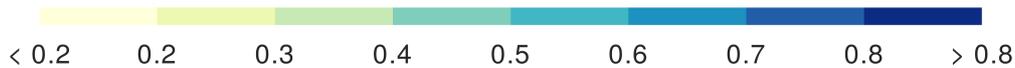
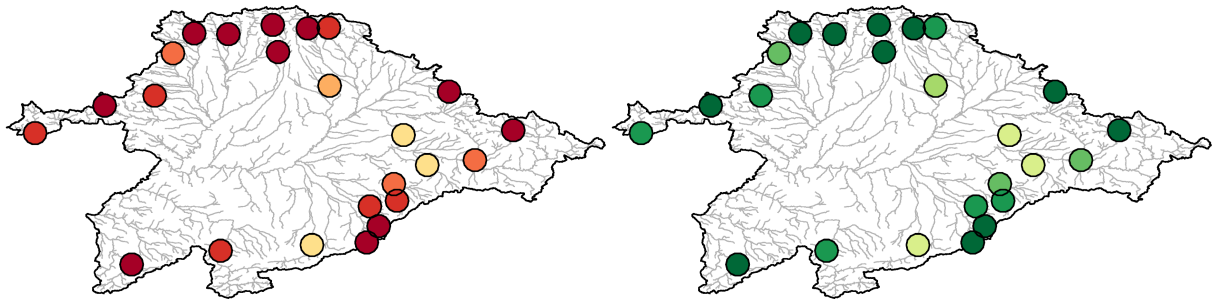
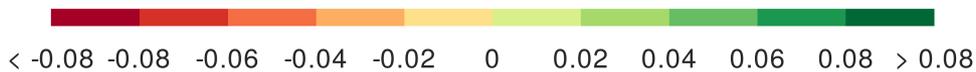

Figure 9

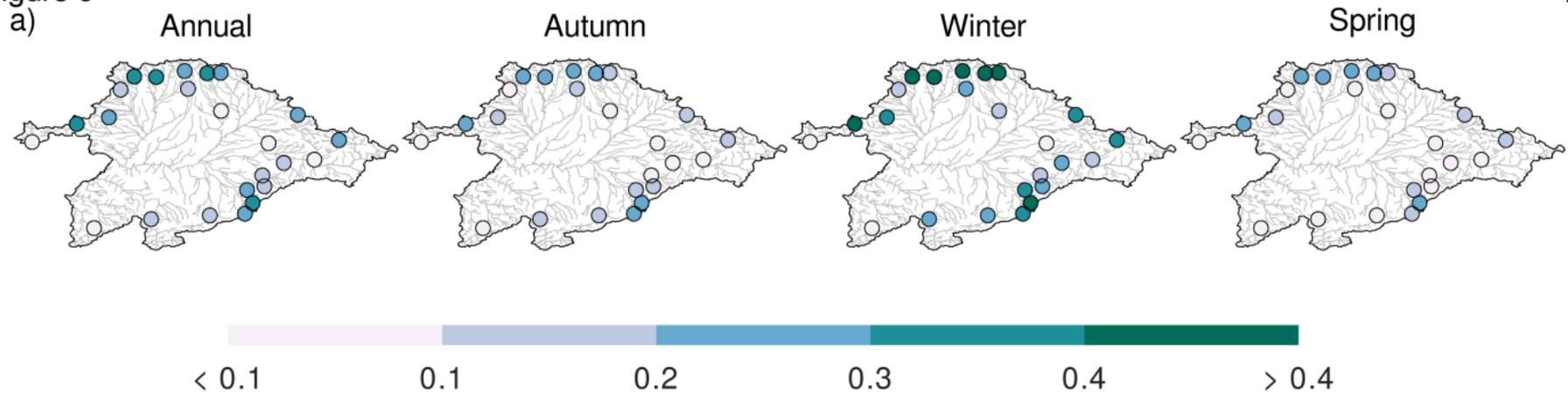

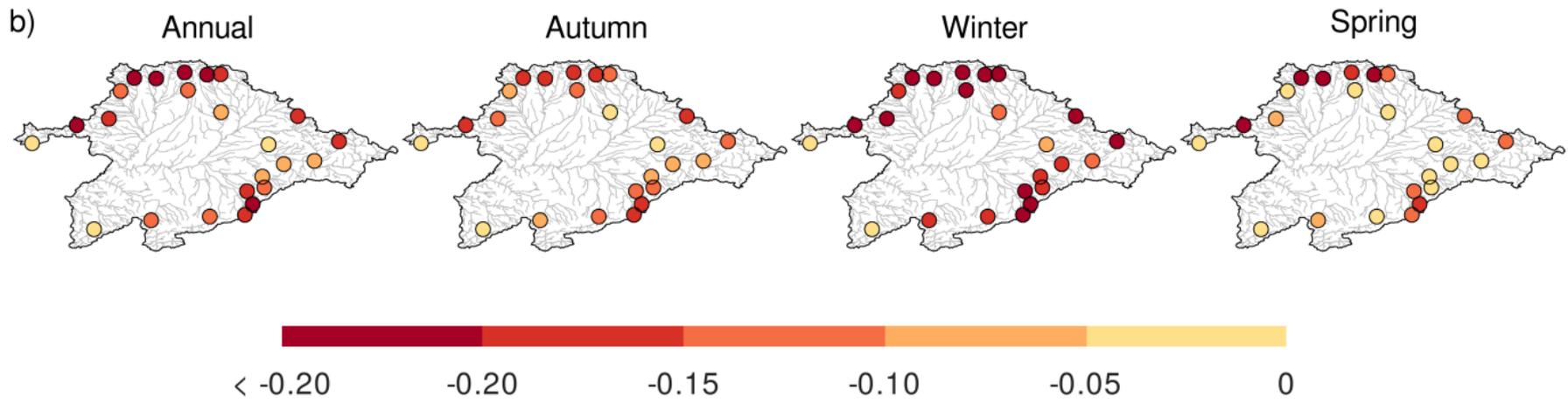

Figure 10

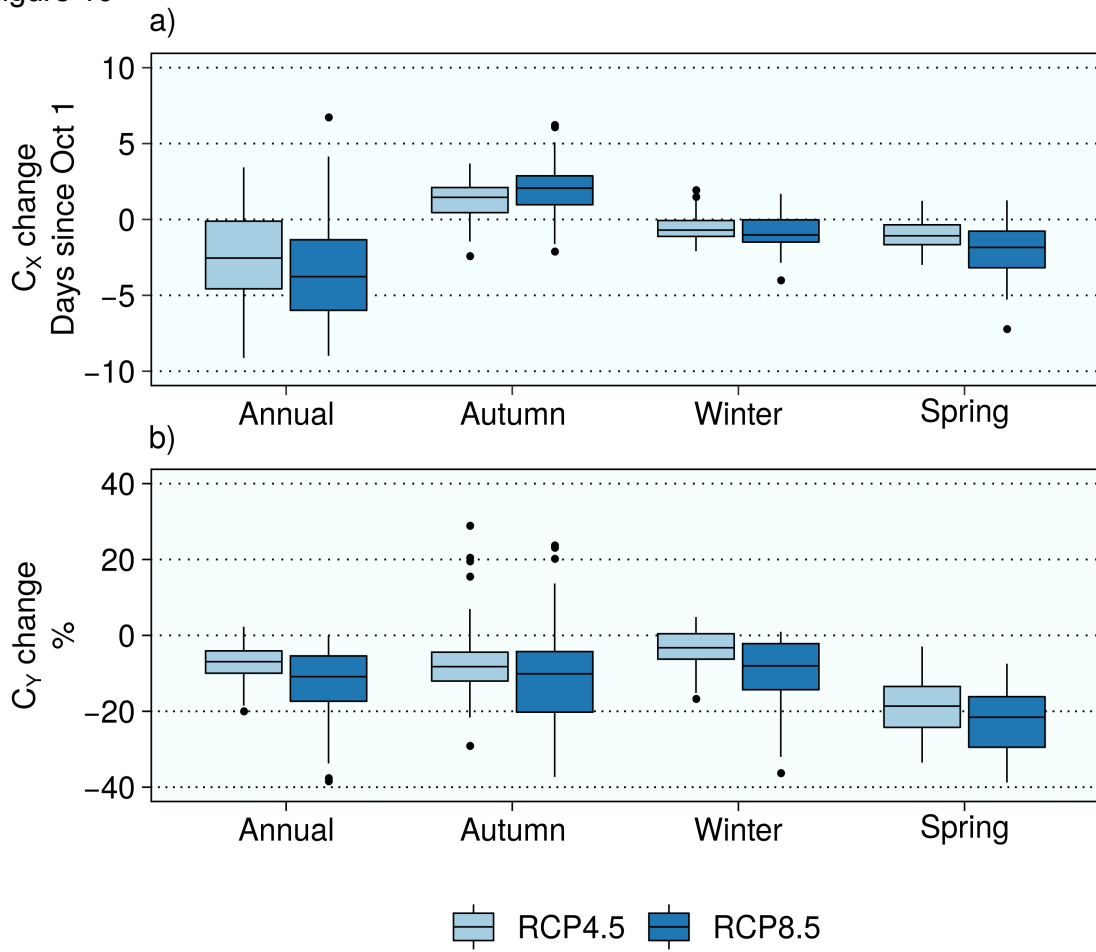